# Differential criterion of a bubble collapse in viscous liquids

Vladislav A. Bogoyavlenskiy*

*Low Temperature Physics Department, Moscow State University, 119899 Moscow, Russia*

(Received 11 January 1999)

The present work is devoted to a model of bubble collapse in a Newtonian viscous liquid caused by an initial bubble wall motion. The obtained bubble dynamics described by an analytic solution significantly depends on the liquid and bubble parameters. The theory gives two types of bubble behavior: collapse and viscous damping. This results in a general collapse condition proposed as the sufficient differential criterion. The suggested criterion is discussed and successfully applied to the analysis of the void and gas bubble collapse. [S1063-651X(99)01207-6]

## I. INTRODUCTION

Formation and collapse of bubbles in liquids are used in many technical applications such as sonochemistry, lithotripsy, ultrasonic cleaning, bubble chambers, and laser surgery [1–4]. Bubble dynamics has been the subject of intensive theoretical and experimental studies since Lord Rayleigh found the well-known analytic solution of this problem for inviscid liquids [5]. The advanced theory of cavitation developed by Plesset gives the differential Rayleigh-Plesset (RP) equation for the bubble radius $R(t)$ [6]. The RP equation describes the dynamics of a spherical void or gas bubble in viscous liquids [7–10] and is also used as a first approximation in more complex problems such as cavitation near solid boundaries [11–14], collapse of asymmetric bubbles [15,16], and sonoluminescence [17–23].

The main difficulties involved in theoretical investigations of the RP equation are that (i) the solutions can be obtained only numerically and (ii) the bubble wall velocity increases to infinity as the bubble collapses. Thus, computer simulations of the bubble motion take a great deal of time and may lead to significant errors in the numerically calculated solutions, especially when the bubble achieves supersonic speeds. Unfortunately, the analytically described bubble dynamics was obtained only for the collapse in inviscid liquids [5].

In this paper we present a way to avoid the above difficulties for viscous liquids. The concept is based on the fact that the RP equation is analytically integrable in the case of the following restrictions: (i) the bubble is void and (ii) the ambient hydrostatic pressure is absent. The imposed restrictions are valid as the bubble collapses because the gas and the ambient pressures are negligible in comparison to the velocity pressure at the bubble wall. The model gives an analytic solution for the bubble radius $R(t)$ and a collapse criterion in the differential form. This differential criterion is considered to be a sufficient condition of the bubble collapse in viscous liquids.

The present paper is organized as follows. In Sec. II the general model of the void bubble collapse in a viscous liquid caused by an initial bubble wall motion is formulated and solved. The subject of Sec. III is the application of the proposed differential collapse criterion to the Rayleigh problem in a viscous liquid and to the collapse of an air bubble in water caused by periodic acoustic pressure.

## II. GENERAL MODEL

### A. Problem formulation

Let us consider a void bubble immersed in an infinite Newtonian viscous liquid. We assume that the bubble is always spherical. Taking into account the symmetry of this problem, we write all the equations in the spherical coordinate system $(r,\varphi,\theta)$ whose origin is at the center of the bubble. Then the liquid motion is governed by the following equations [1]:

$$\frac{\partial \sigma_{rr}}{\partial r} + \frac{2\sigma_{rr} - \sigma_{\theta\theta} - \sigma_{\varphi\varphi}}{r} = \rho\left(\frac{\partial v_r}{\partial t} + \frac{1}{2}\frac{\partial v_r^2}{\partial r}\right), \quad (1)$$

$$\frac{\partial v_r}{\partial r} + 2\frac{v_r}{r} = 0, \quad (2)$$

$$\sigma_{rr} = -p + 2\mu\frac{\partial v_r}{\partial r}, \quad \sigma_{\theta\theta} = \sigma_{\varphi\varphi} = -p + 2\mu\frac{v_r}{r}, \quad (3)$$

where $r$ is the radial coordinate, $v_r$ is the radial liquid velocity, $\rho$=const is the liquid density, $\sigma$ is the stress tensor, $p$ is the hydrostatic pressure, and $\mu$=const is the shear viscosity. For this set of equations to be complete, we add the initial and the boundary conditions. Let us assume that the ambient pressure and the surface tension are negligible. These restrictions result in the following conditions on the bubble surface and at infinity:

$$\sigma_{rr}\{r=R(t)\}=0, \quad \sigma_{rr}\{r=\infty\}=0. \quad (4)$$

The initial conditions are chosen to be nonstandard. Usually the initial bubble wall motion is ignored, but in this model the bubble is considered to have a radial velocity $V_0$:

$$R\{t=0\}=R_0, \quad \frac{dR}{dt}\{t=0\}=-V_0. \quad (5)$$

---

*Electronic address: bogoyavlenskiy@usa.net

The system of Eqs. (1)–(5) completely describes the model. To find the solution, let us use the method proposed by Rayleigh [5]. According to the incompressibility condition given by Eq. (2), the radial liquid velocity can be written as

$$v_r = \frac{dR}{dt}\left(\frac{R}{r}\right)^2. \quad (6)$$

After the substitution of Eqs. (3) and (6) into Eq. (1) and its subsequent integration in the range $(R,\infty)$ we write the expression

$$\int_R^\infty \left(\frac{\partial \sigma_{rr}}{\partial r} - \frac{12\mu R^2}{r^4}\frac{dR}{dt}\right)dr = \rho \int_R^\infty \left[\frac{R^2}{r^2}\frac{d^2R}{dt^2} + \frac{2R}{r^2}\left(\frac{dR}{dt}\right)^2 - \frac{2R^4}{r^5}\left(\frac{dR}{dt}\right)^2\right]dr. \quad (7)$$

Taking into account Eq. (4), we obtain the modified RP equation where the initial conditions are given by Eq. (5):

$$R\frac{d^2R}{dt^2} + \frac{3}{2}\left(\frac{dR}{dt}\right)^2 + \frac{4\mu}{\rho R}\frac{dR}{dt} = 0. \quad (8)$$

### B. Solution and analysis

Let us define the following dimensionless variables and constants:

$$\tilde{R} \equiv \frac{R}{R_0}, \quad \tilde{t} \equiv t\frac{V_0}{R_0}, \quad \tilde{\mu} \equiv \frac{\mu}{\rho R_0 V_0}, \quad \alpha \equiv \frac{1}{8\tilde{\mu}} - 1. \quad (9)$$

Here $\tilde{R}$, $\tilde{t}$, and $\tilde{\mu}$ are the dimensionless bubble radius, time, and viscosity, respectively. Then Eqs. (8) and (5) can be represented as

$$\tilde{R}\frac{d^2\tilde{R}}{d\tilde{t}^2} + \frac{3}{2}\left(\frac{d\tilde{R}}{d\tilde{t}}\right)^2 + \frac{4\tilde{\mu}}{\tilde{R}}\frac{d\tilde{R}}{d\tilde{t}} = 0,$$

$$\tilde{R}\{\tilde{t}=0\}=1, \quad \frac{d\tilde{R}}{d\tilde{t}}\{\tilde{t}=0\}=-1. \quad (10)$$

The differential equation (10) is integrable and the obtained analytic solution is the following:

$$\tilde{t} = 2(\alpha+1)\left(\frac{1}{4}(1-\tilde{R}^2) - \frac{\alpha}{3}(1-\tilde{R}^{3/2}) + \frac{\alpha^2}{2}(1-\tilde{R}) - \alpha^3(1-\tilde{R}^{1/2}) - \alpha^4\ln\frac{\alpha+\tilde{R}^{1/2}}{\alpha+1}\right). \quad (11)$$

The most illustrative way to discuss the bubble dynamics is by analysis of the kinetic energy accumulated by the liquid near the bubble wall. The dimensionless expression of this energy is given by the relation

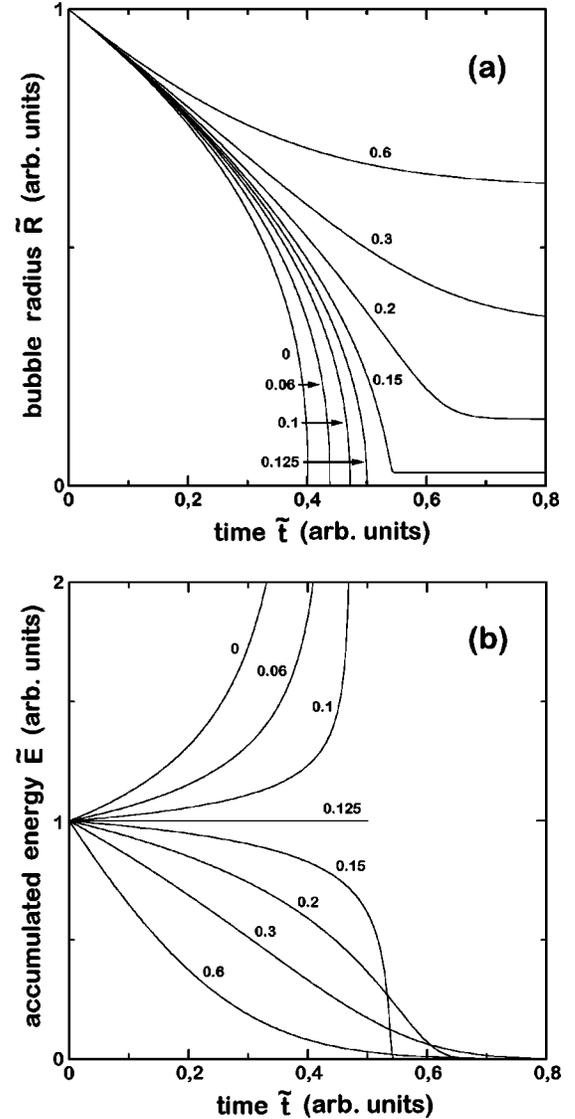

FIG. 1. Bubble radius $\tilde{R}$ (a) and accumulated energy $\tilde{E}$ (b) as functions of time $\tilde{t}$. Values of parameter $\tilde{\mu}$ are shown at curves.

$$\tilde{E} \equiv \tilde{R}^2\left(\frac{d\tilde{R}}{d\tilde{t}}\right)^2$$

$$= \frac{\tilde{R}^2}{(\alpha+1)^2\left(-\tilde{R}+\alpha\tilde{R}^{1/2}-\alpha^2+\alpha^3\tilde{R}^{-1/2}-\frac{\alpha^4\tilde{R}^{-1/2}}{\alpha+\tilde{R}^{1/2}}\right)^2}. \quad (12)$$

The bubble collapse corresponds to the condition $\tilde{E}\to\infty$.

The overall picture of the bubble behavior given by Eqs. (11) and (12) is summarized by Fig. 1, which shows the time dependence of the bubble radius $\tilde{R}(\tilde{t})$ [Fig. 1(a)] and the accumulated energy $\tilde{E}(\tilde{t})$ [Fig. 1(b)]. The behavior of the curves significantly depends on the value of the dimensionless viscosity $\tilde{\mu}$. Analysis of Eq. (11) shows that the bubble radius $\tilde{R}(\tilde{t})$ decreases to zero only if $\tilde{\mu}<\frac{1}{8}$. In this case, the accumulated energy $\tilde{E}(\tilde{t})$ increases to infinity. That is, the bubble collapse takes place:

$$\lim_{\widetilde{R}\to 0}\widetilde{E}=\frac{\alpha^2}{(\alpha+1)^2}\lim_{\widetilde{R}\to 0}\frac{1}{\widetilde{R}}=\infty. \quad (13)$$

Analysis of the bubble motion at $\widetilde{R}(\widetilde{t})\to 0$ gives the following approximation formula:

$$\widetilde{R}=\left(\frac{5\alpha}{2(\alpha+1)}(\widetilde{t}-\widetilde{t}_C)\right)^{2/5}, \quad (14)$$

$$\widetilde{t}_C\equiv\widetilde{t}\{\widetilde{R}=0\}=2(\alpha+1)\left(\frac{1}{4}-\frac{\alpha}{3}+\frac{\alpha^2}{2}-\alpha^3-\alpha^4\ln\frac{\alpha}{\alpha+1}\right). \quad (15)$$

Here $\widetilde{t}_C$ is the collapse time, which varies from 0.4 ($\widetilde{\mu}\to 0$) to 0.5 ($\widetilde{\mu}=\frac{1}{8}$). The bubble dynamics described by Eqs. (14) and (15) is basically similar to one obtained by Rayleigh [5].

The other case $\widetilde{\mu}>\frac{1}{8}$ corresponds to the viscous damping. The bubble radius $\widetilde{R}(\widetilde{t})$ smoothly decreases to the equilibrium value $\widetilde{R}_{eq}$ and the accumulated energy $\widetilde{E}(\widetilde{t})$ descends to zero:

$$\widetilde{R}_{eq}=\alpha^2=\left(\frac{1}{8\widetilde{\mu}}-1\right)^2, \quad (16)$$

$$\lim_{\widetilde{R}\to\widetilde{R}_{eq}}\widetilde{E}=\frac{1}{\alpha^2(\alpha+1)^2}\lim_{\widetilde{R}\to\widetilde{R}_{eq}}(\widetilde{R}^{1/2}+\alpha)^2=0. \quad (17)$$

### III. APPLICATIONS OF THEORY

The condition of the bubble collapse $\widetilde{\mu}<\frac{1}{8}$ is the main result obtained in the preceding section. Let us rewrite this inequality as

$$\left(-\frac{dR}{dt}\{R=R_0\}\right)\frac{\rho R_0}{8\mu}>1. \quad (18)$$

We should emphasize the special features of the model presented that result from the boundary conditions. The above inequality contains only one variable $R(t)$ and two liquid constants $\rho$ and $\mu$. Moreover, Eq. (18) is a local, differential condition, which means there is no information about the preceding bubble motion. The condition (18) is insensitive to the substitution $R_0\leftrightarrow R(t)$. Thus, the above collapse condition can be represented as

$$\widetilde{c}(t)>1, \quad \widetilde{c}(t)\equiv\left(-\frac{dR(t)}{dt}\right)\frac{\rho R(t)}{8\mu}, \quad (19)$$

where $\widetilde{c}(t)$ is the dimensionless *collapse variable*.

The physics of the differential condition (19) is quite simple. Let us focus on a bubble motion governed by the system of equations (1)–(3) (see Sec. II) in the presence of an ambient hydrostatic pressure and a gas pressure inside the bubble. Instead of the curve $R(t)$, the behavior of the curve $\widetilde{c}(t)$ is analyzed. If the value of $\widetilde{c}(t)$ achieves the number one, the collapse takes place. This is the sufficient condition for a void bubble, since the ambient hydrostatic pressure additionally forces the bubble to the collapse. The case of the gas bubble is more complicated because the gas pressure slows down the bubble wall motion. However, in most cases the gas pressure is negligible in comparison to the velocity pressure as the criterion (19) is realized. Two applications of the proposed criterion are presented below.

### A. Rayleigh's problem in a viscous liquid

The Rayleigh problem is the study of void bubble motion in a liquid caused by a constant ambient pressure [5]. In this case the boundary and initial conditions are transformed from Eqs. (4) and (5) to the following:

$$\sigma_{rr}\{r=R\}=0, \quad \sigma_{rr}\{r=\infty\}=p_0, \quad (20)$$

$$R\{t=0\}=R_0, \quad \frac{dR}{dt}\{t=0\}=0, \quad (21)$$

where $p_0=$const is the ambient hydrostatic pressure. After repeating the sequence of procedures described in Sec. II, we obtain the RP equation [6]

$$R\frac{d^2R}{dt^2}+\frac{3}{2}\left(\frac{dR}{dt}\right)^2+\frac{4\mu}{\rho R}\frac{dR}{dt}=-\frac{p_0}{\rho}. \quad (22)$$

The problem has the well-known analytic solution for inviscid liquids found by Rayleigh in 1917 [5]. In the case of a Newtonian viscous liquid, the numerical solution was obtained by Zababakhin [10]. The computer simulations of the bubble motion show two types of bubble behavior: a collapse and a smooth decrease, where the collapse condition can be written as

$$\widetilde{\mu}_p<0.119, \quad \widetilde{\mu}_p\equiv\frac{\mu}{R_0\sqrt{\rho p_0}}. \quad (23)$$

To illustrate the advantages of the differential criterion (19) obtained, the time dependence of the bubble radius $R(t)$ and the collapse variable $\widetilde{c}(t)$ at various values of $\widetilde{\mu}_p$ are presented in Fig. 2. Calculation of the bubble radius $R(t)$ shows that the critical value of $\widetilde{\mu}_p$ corresponding to the collapse criterion lies within the interval 0.1–0.12 [see Fig. 2(a)]. More precise estimates are hampered by instabilities and errors in the numerical procedure as $R(t)\to 0$.

We propose finding the collapse condition by analyzing the collapse variable $\widetilde{c}(t)$ [see Fig. 2(b)]. When the maximum of the curve $\widetilde{c}(t)$ is less than 1, the curve corresponds to the viscous damping of the bubble wall motion. The bubble collapse is realized when the curve $\widetilde{c}(t)$ exceeds 1. The $\widetilde{c}(t)$ analysis significantly reduces the numerical error in comparison to the $R(t)$ analysis. This results in the most precise collapse criterion:

$$\widetilde{\mu}_p<(0.114\,63\pm 0.000\,01). \quad (24)$$

### B. Collapse of an air bubble in water caused by sound

The condition we consider is an air bubble in water subjected to a periodic spherical sound wave of ultrasonic fre-

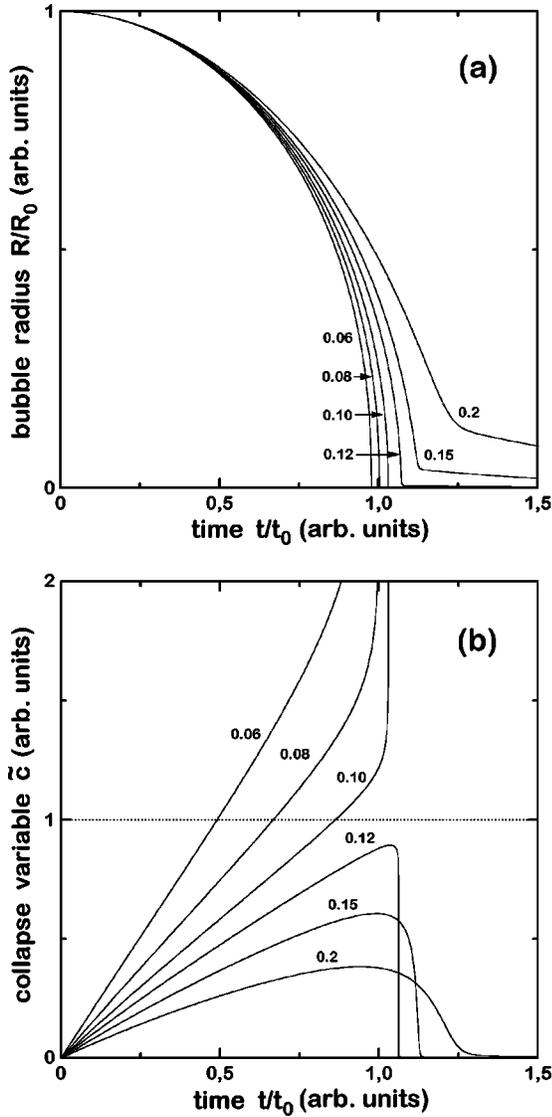

FIG. 2. Bubble radius $R/R_0$ (a) and collapse variable $\tilde{c}$ (b) as functions of time $t/t_0$ where $t_0 \equiv R_0\sqrt{\rho/p_0}$. Values of parameter $\tilde{\mu}_p$ are shown at curves. The dotted line corresponds to the collapse criterion $\tilde{c}(t) = 1$.

quency [7,8]. Assuming the symmetry is spherical, the bubble radius $R(t)$ obeys the following equations:

$$R\frac{d^2R}{dt^2} + \frac{3}{2}\left(\frac{dR}{dt}\right)^2 + \frac{4\mu}{\rho R}\frac{dR}{dt}$$
$$= \frac{p_g - p_a - p_0}{\rho} + \frac{R}{\rho c}\frac{d}{dt}(p_g - p_a), \quad (25)$$

$$R\{t=0\} = R_0, \quad \frac{dR}{dt}\{t=0\} = 0. \quad (26)$$

Here $R_0$ is the equilibrium bubble radius, $\rho$ is the water density, $\mu$ is the shear viscosity of the water, $c$ is the speed of sound in water, and $p_0 = $ const is the ambient hydrostatic pressure. The acoustic pressure $p_a$ is considered to be periodic:

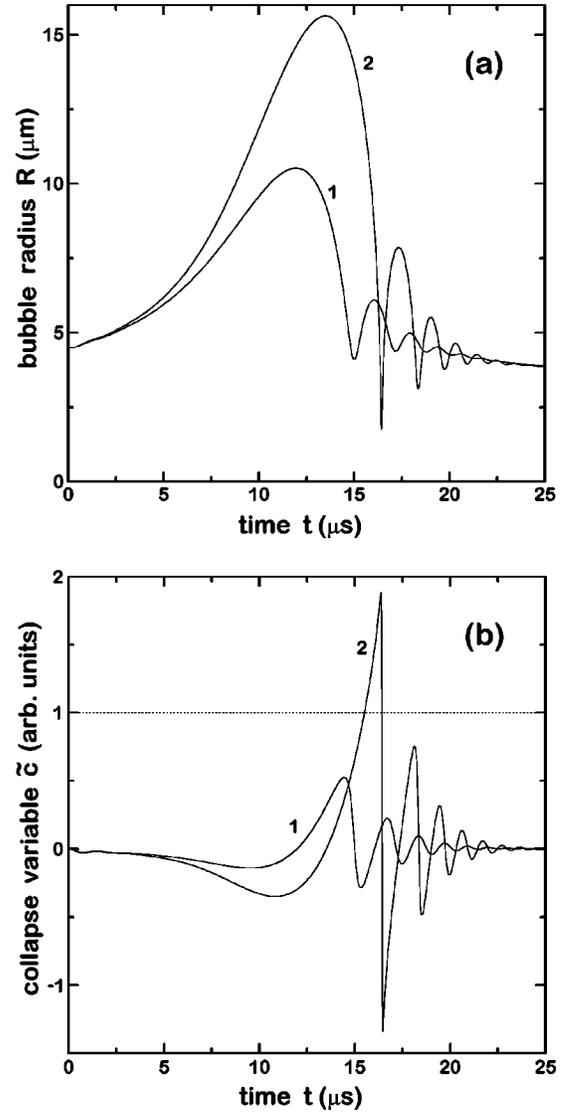

FIG. 3. Bubble radius $R$ (a) and collapse variable $\tilde{c}$ (b) as functions of time $t$ during one acoustic period. These are results for the following parameters: $\rho = 1.0$ g/cm, $c = 1481$ m/s, $\mu = 0.07$ g/(cm $\cdot$ s), $R_0 = 4.5$ $\mu$m, $R_0/a = 8.5$, $\gamma = 1.4$, $\omega = 26.5$ kHz, $p_0 = 1.0$ atm. Curves 1 and 2 correspond to $p_a^0 = 0.98$ atm and $p_a^0 = 1.06$ atm, respectively. The dotted line corresponds to the collapse criterion $\tilde{c}(t) = 1$.

$$p_a = -p_a^0 \sin 2\pi\omega t, \quad (27)$$

where $p_a^0$ and $\omega$ are the amplitude and the frequency of the sound wave, respectively. Assuming adiabatic conditions inside the bubble, the gas pressure $p_g$ follows from the van der Waals equation,

$$p_g = p_0\left(\frac{R_0^3 - a^3}{R^3 - a^3}\right)^\gamma, \quad (28)$$

where $a$ is the van der Waals hard core and $\gamma$ is the ratio of specific heats.

The set of equations (25)–(28) describes the nonlinear bubble oscillations that can concentrate the average sound energy by over 12 orders of magnitude [19]. During the acoustic cycle the bubble absorbs the energy from the sound field and its radius expands from the equilibrium value $R_0$ to a maximum value. The subsequent compressional part of the sound field causes the bubble to collapse. Heating of the bubble surface caused by the compression may lead to the emission of a pulse of light as the bubble approaches a minimum radius. This phenomenon is known as sonoluminescence [3,4].

Let us illustrate the application of the collapse criterion (19) to this problem. The calculations of the bubble motion are performed for an air bubble in water where all the values of the parameters in Eqs. (25)–(28) are taken from Refs. [20–22]. For this problem the bubble behavior is basically governed by the amplitude of the acoustic pressure $p_a^0$ and by the equilibrium bubble radius $R_0$. Let us consider that $R_0$ = const. Therefore, the only variable of the problem is the amplitude of the sound wave $p_a^0$.

The calculated bubble radius $R(t)$ and the criterion variable $\tilde{c}(t)$ at various values of $p_a^0$ are presented in Fig. 3. The figure shows that the features of the bubble oscillations are determined by the behavior of $\tilde{c}(t)$. The viscous damping [curve 1 in Fig. 3(a)] corresponds to the inequality $\tilde{c}(t)<1$ during the acoustic cycle [curve 1 in Fig. 3(b)]. As a result, the sound wave energy dissipates only by the shear viscosity of the water. Thus, the increase of the air temperature inside the bubble is negligible. The bubble behavior significantly changes when $\tilde{c}(t)$ exceeds 1 [curve 2 in Fig. 3(b)]. In this case the acoustic energy is focused on the bubble and compresses the air within it to high pressures and temperatures [curve 2 in Fig. 3(a)].

It is significant that the bubble collapse is not the sufficient condition for sonoluminescense. The emission of light occurs only when the energy of the sound field achieves a critical value. For this set of parameters the focused energy drastically increases with the increase of $p_a^0$. The edge of sonoluminescence corresponds to the value of $p_a^0 \sim 1.2$ atm [20].

## ACKNOWLEDGMENTS

I would like to thank Dr. N. A. Chernova and Dr. D. V. Georgievskii for useful discussions. I would also like to acknowledge Ms. Tana Mierau for helpful comments.